\documentclass{article}
\usepackage[preprint]{spconf}
\usepackage{amsmath,graphicx,hyperref}

\usepackage{amsmath, amssymb, bm}

\usepackage{multirow}
\usepackage{colortbl}
\usepackage{xcolor}
\usepackage{booktabs} 
\usepackage{caption}  

\title{IF-D: A High-Frequency, General-Purpose Inertial Foundation Dataset for Self-Supervised Learning}
\name{Patrick Ferreira, Paula Costa}

\address{Faculdade de Engenharia Elétrica e de Computação (FEEC)\\
Universidade Estadual de Campinas (UNICAMP)\\
Campinas, Brazil}

\begin{document}
%
\maketitle

\copyrightnotice{\copyright\ 2026 IEEE. Personal use of this material is permitted. Permission from IEEE must be obtained for all other uses, in any current or future media. 
Including reprinting/republishing this material for advertising or promotional purposes, creating new collective works, for resale or redistribution to servers or lists, or reuse of any copyrighted component of this work in other works. Submitted to IEEE ICASSP 2026.}

\begin{abstract}
We present IF-D, a large-scale inertial dataset designed to enable self-supervised and foundational learning for IMU time series. IF-D comprises continuous, long-duration multichannel recordings (accelerometer, gyroscope, magnetometer) sampled at 200Hz using a UM7 IMU mounted inside a 3D-printed spherical enclosure that promotes diverse, free rotations during vehicle traversal. The collection spans approximately 135 minutes of recording, yielding around 1.6 million samples across nine sensor channels. We describe the data acquisition setup, preprocessing, and calibration procedures (six-orientation accelerometer calibration, stationary gyroscope bias estimation, and ellipsoid fitting for magnetometer hard-/soft-iron correction), and provide quantitative calibration results. IF-D is designed to mitigate platform-specific motion bias and expose models to both physical dynamics and typical measurement noise, thereby facilitating robust representation learning and downstream tasks such as event detection, motion mode recognition, and inertial navigation.
\end{abstract}

\section{Introduction}
\label{sec:intro}

Inertial sensors (accelerometers, gyroscopes, magnetometers) are central to robotics, navigation, wearable sensing and industrial systems due to their capacity to provide motion information independent of external references \cite{ElSheimy2020,Gao2022}. However, raw IMU time series pose specific challenges: they typically lack clear periodicity, present high-dimensional coupled channels, and are affected by significant noise and bias that degrade naive integration and linear filtering. These characteristics limit the applicability of classical time-series assumptions (stationarity, separability) and motivate methods that capture subtle temporal and cross-channel dependencies.

Motivated by the training paradigm of large language models—which learn both real-world structure and recurring artifacts from massive, minimally preprocessed corpora—we propose to collect a large, diverse inertial corpus to enable analogous foundational learning for IMU data. To this end we introduce the Inertial Foundation Dataset (IF-D): long, high-frequency (200Hz) multichannel sequences captured with a UM7 IMU mounted inside a 3D-printed sphere to maximize motion diversity and reduce platform-specific bias. We expect IF-D to support large-scale self-supervised pre-training and the development of robust models that learn both physical patterns and characteristic measurement errors, as suggested by recent foundational-model efforts in other temporal domains \cite{Zhao2025TartanIMU,Wieland2025}.

\section{Related Work}
\label{sec:related}

Transformer-based architectures and large-scale pre-training have advanced time-series forecasting by capturing long-range dependencies. Representative models include Informer, Autoformer and FEDformer for efficient attention and decomposition \cite{Zhou2021Informer,Wu2021Autoformer,Zhou2022FEDformer}; PatchTST and SparseTSF for computationally efficient local/ sparse modeling \cite{Nie2023PatchTST,Lin2024SparseTSF}; and foundational/time-scale efforts such as TimesFM and structural approaches like TimesNet \cite{Das2024TimesFM,Wu2023TimesNet}. These methods have been validated mainly on series with clear trends or seasonality (weather, traffic, finance), where stable temporal regularities favor attention-based extrapolation.

Conversely, publicly available inertial datasets are specialized and limited in scope (sampling rate, motion regimes or duration), e.g., UCI HAR \cite{Anguita2013}, RoNIN \cite{Yan2020RoNIN}, EuRoC MAV \cite{Burri2016EuRoC}, TUM VI \cite{Schubert2018TUMVI}, and KITTI \cite{Geiger2013KITTI}. Each covers a subspace of motions (walking, drone flights, vehicle trajectories) and thus embeds application-specific bias that impedes generalization. This fragmented landscape motivates IF-D, which aims to provide long, high-fidelity, diverse multichannel inertial data suitable for self-supervised and Transformer-based pre-training and for bridging the gap between existing time-series foundations and inertial sensing applications.

\section{Materials and Methods}
\label{sec:materials}

The dataset was collected using the \textbf{IMU UM7} model from CH Robotics, which features a sampling rate of 200 Hz and provides three degrees of freedom in the measurement of linear acceleration, angular velocity, and magnetic field. The platform used for data acquisition was the NVIDIA \textbf{Jetson Nano}, chosen for its portability and robust embedded hardware.

\begin{figure}[thb]
  \centering
  \includegraphics[width=0.2\linewidth]{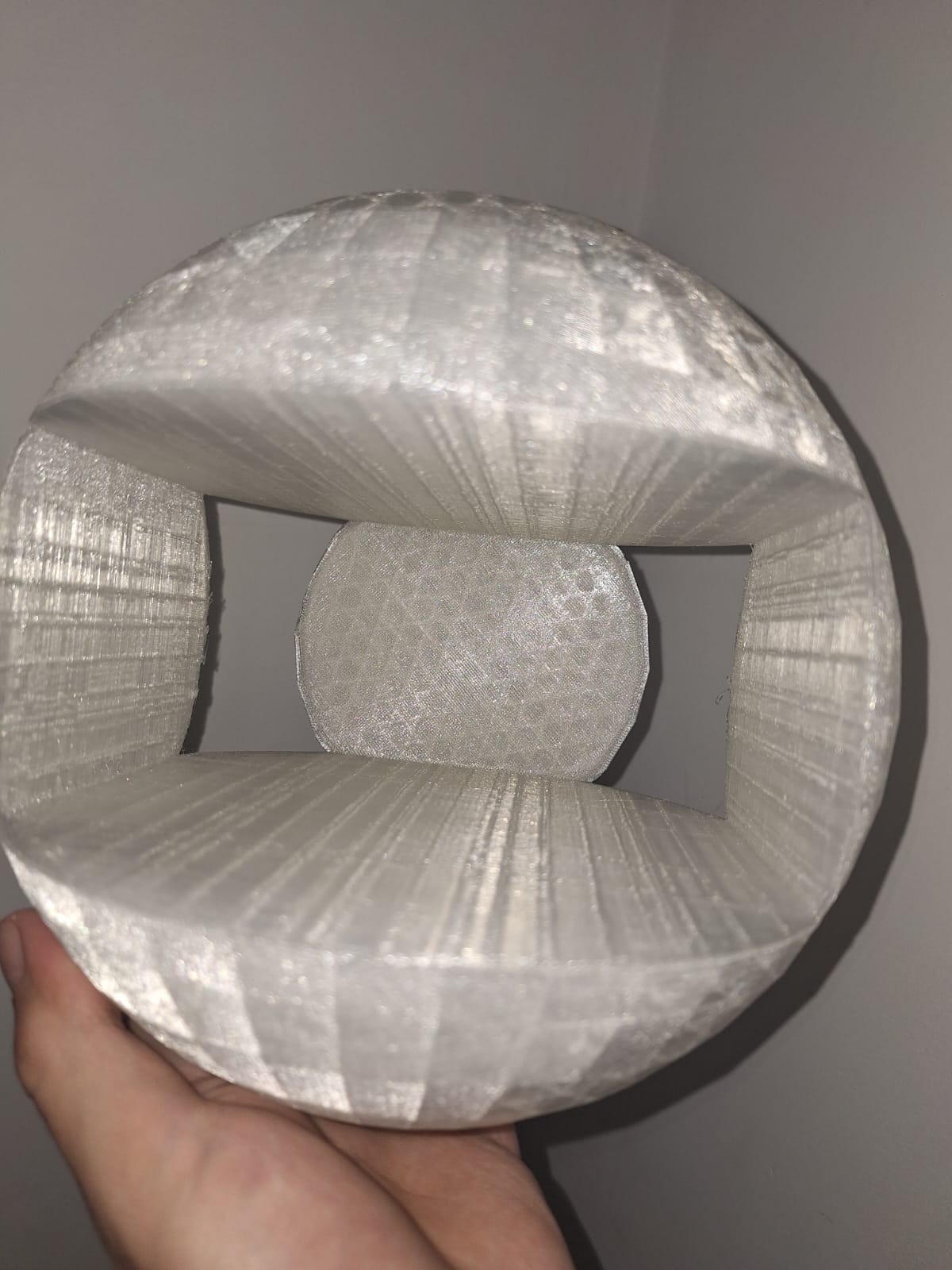}
  \includegraphics[width=0.2\linewidth]{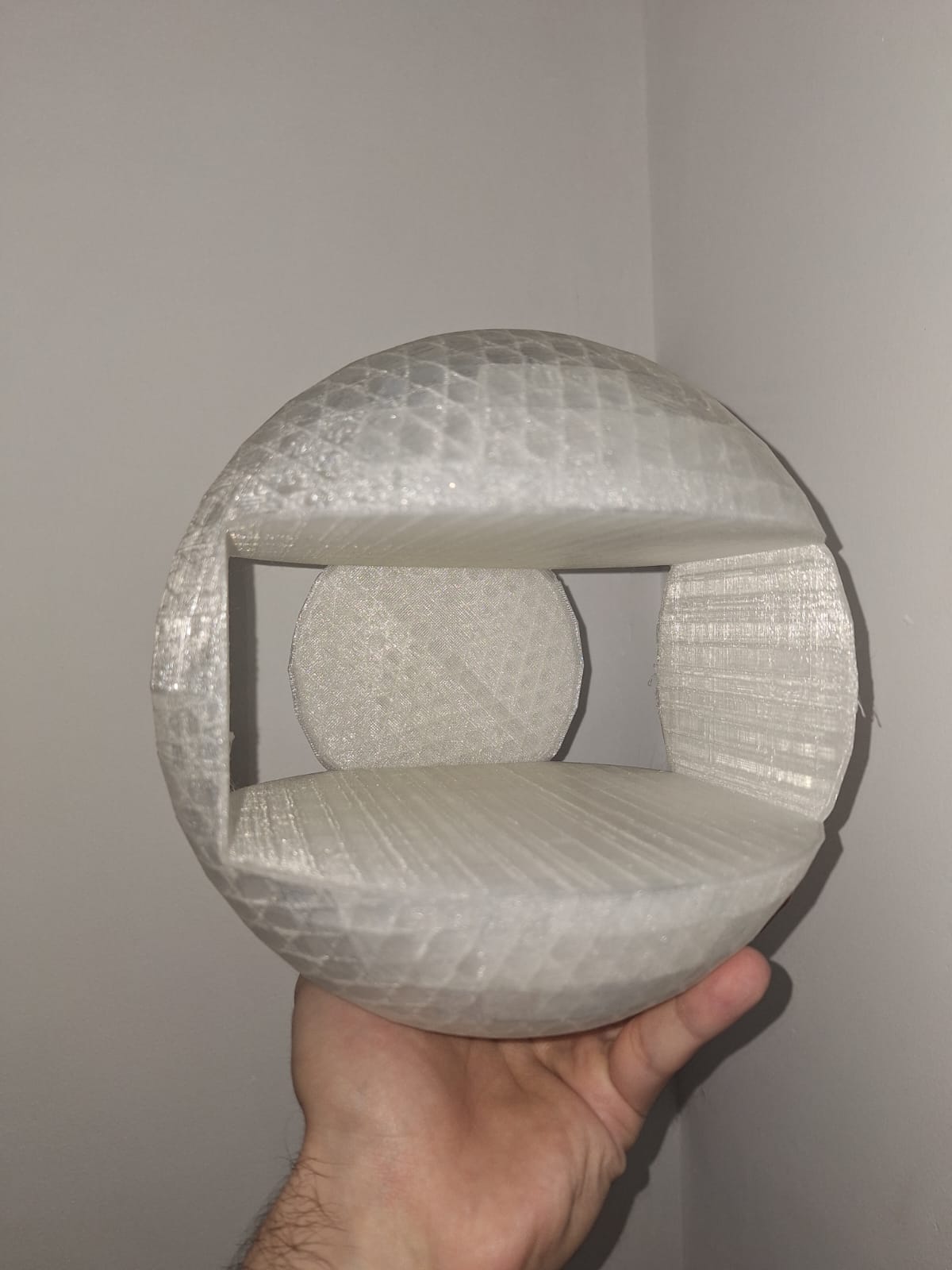}
  \includegraphics[width=0.2\linewidth]{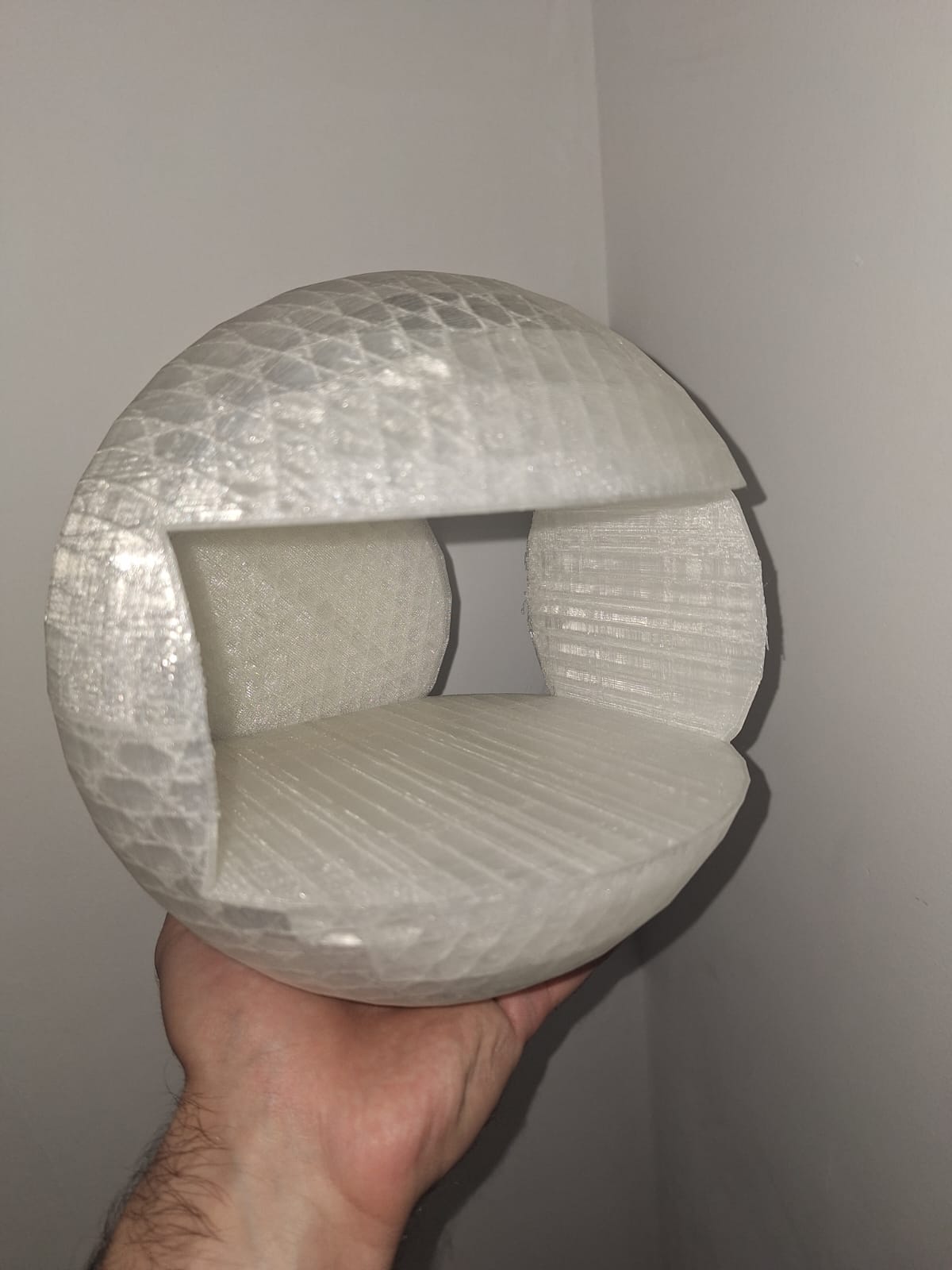}
  \includegraphics[width=0.2\linewidth]{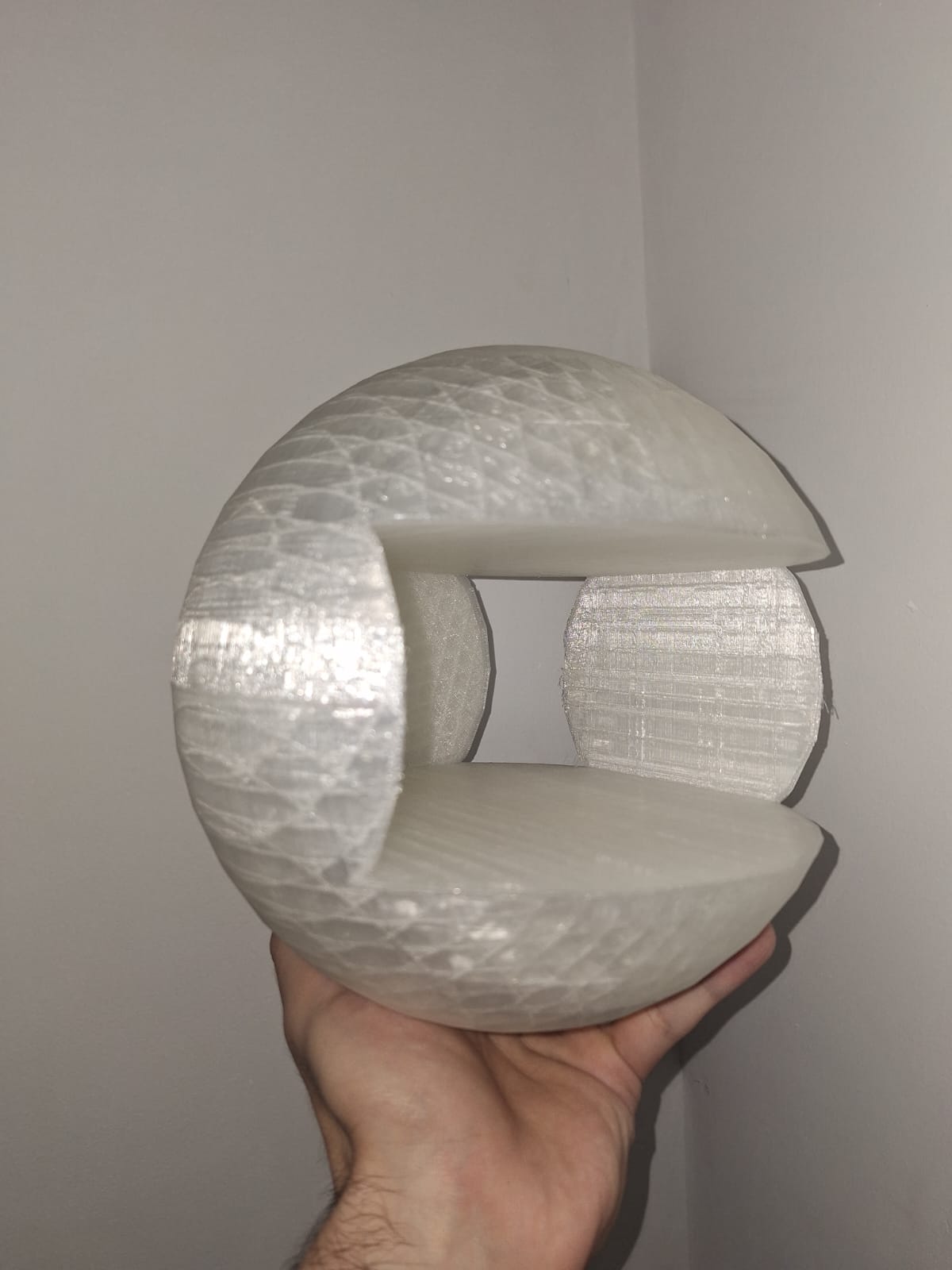}
  \\
  \includegraphics[width=0.2\linewidth]{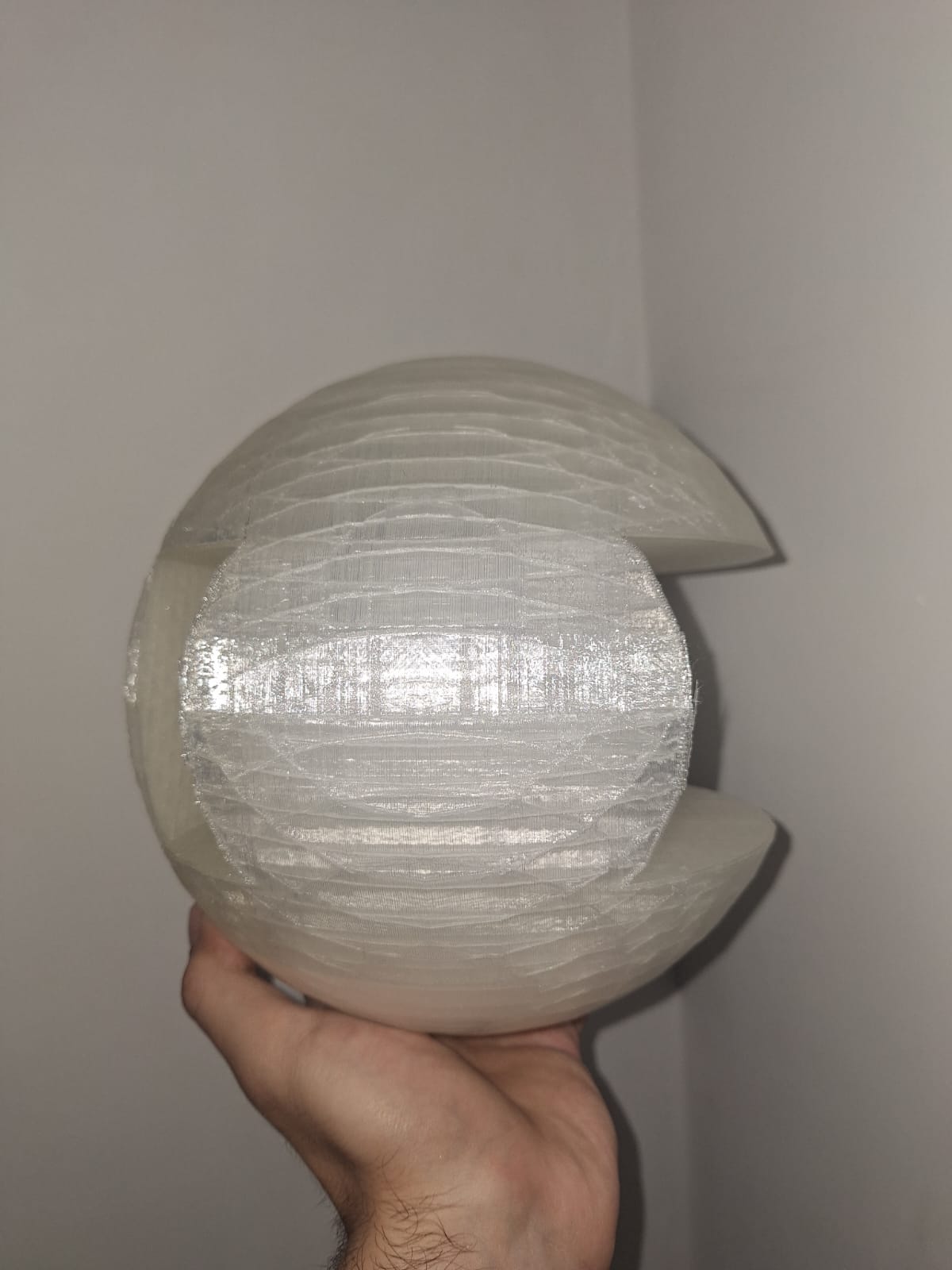}
  \includegraphics[width=0.2\linewidth]{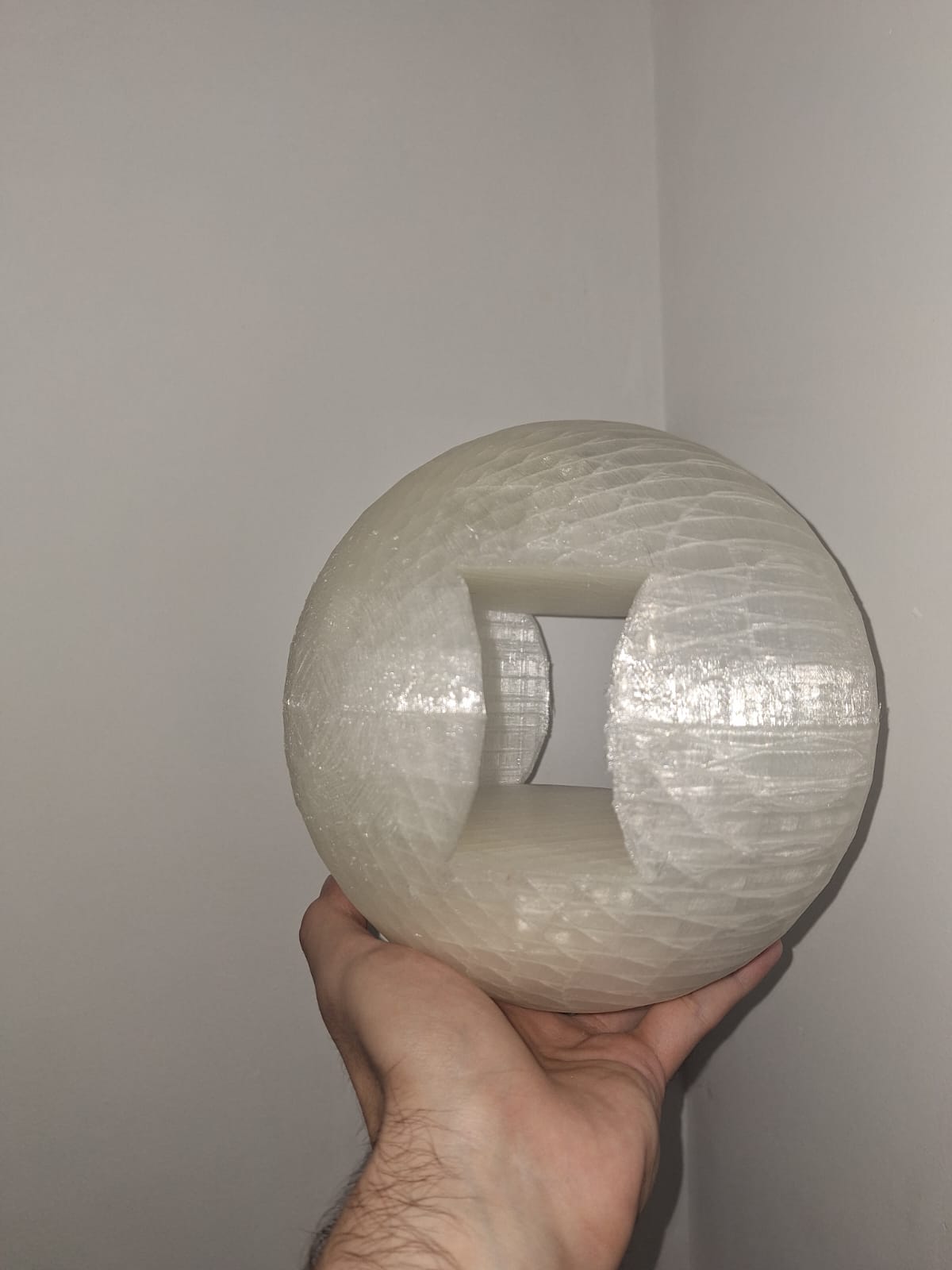}
  \includegraphics[width=0.2\linewidth]{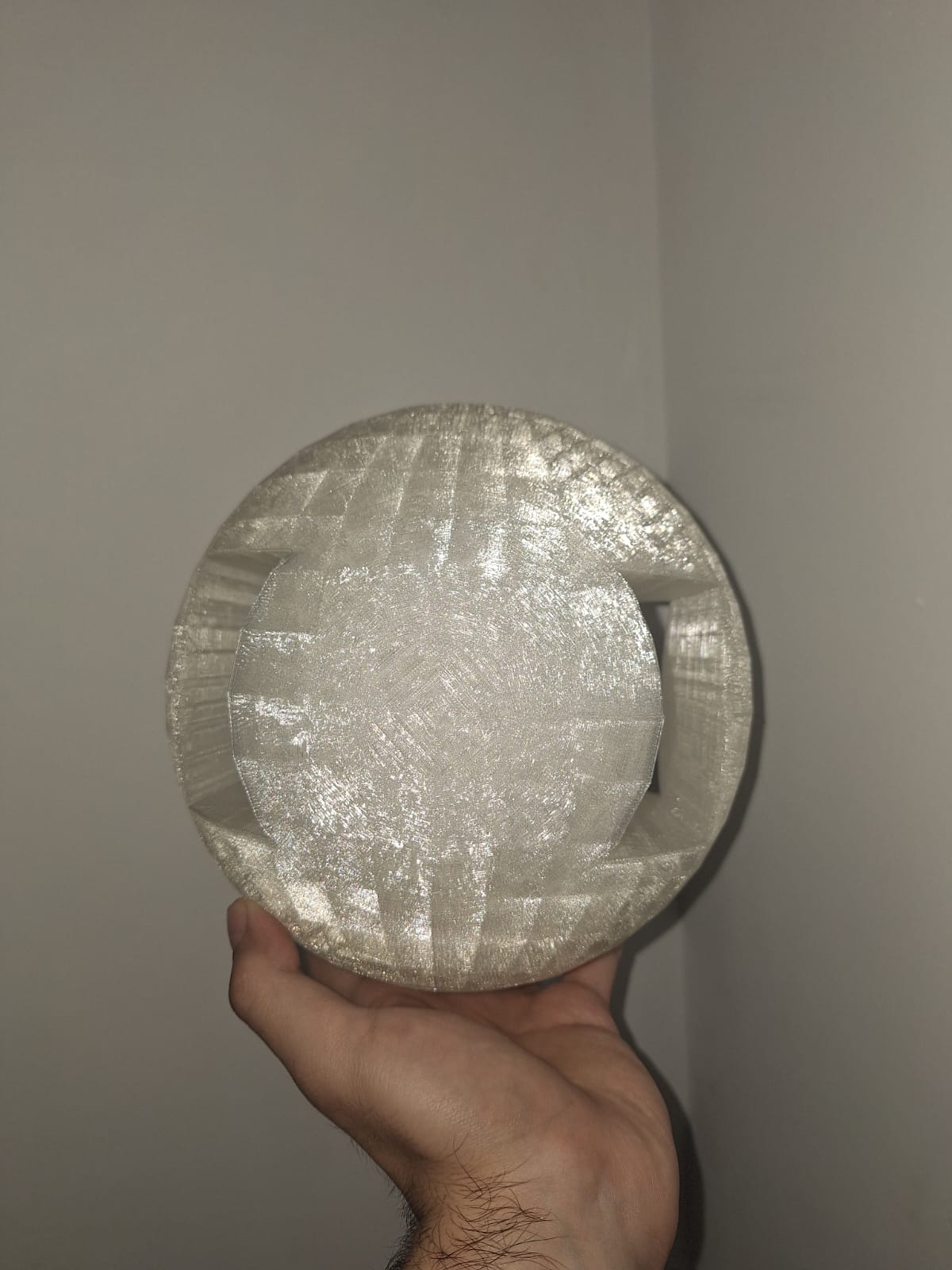}
  \includegraphics[width=0.2\linewidth]{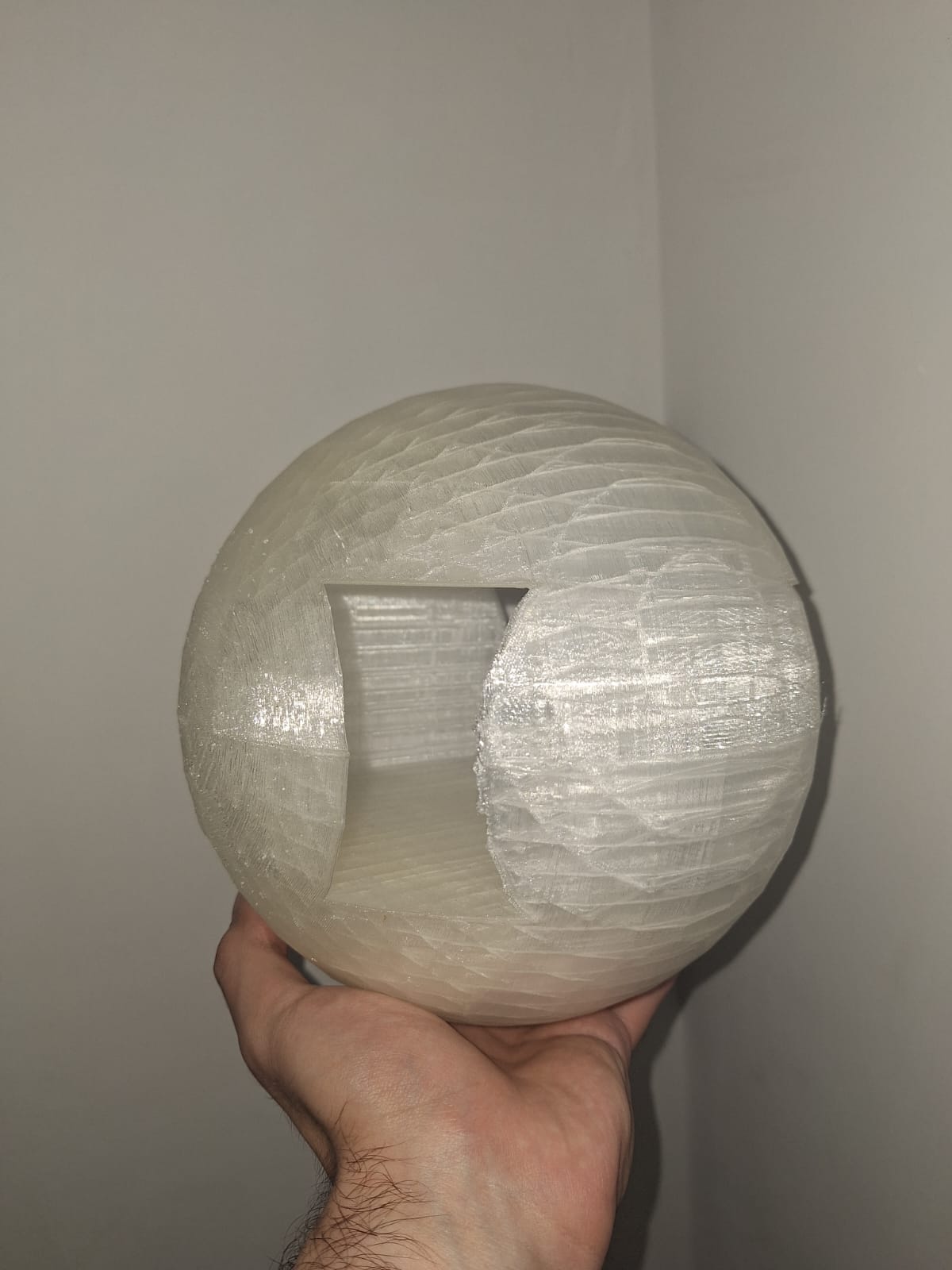}
  \caption{Multiple views of the 3D-printed spherical model enabling free movement during dataset collection.}
  \label{fig:impressao-3d}
\end{figure} 

Inertial datasets are subject to biases in data collection due to the motion patterns of the vehicle or platform used for acquisition \cite{Shurin2022}. To maximize the range of possible trajectories, a spherical structure for data collection was devised, 3D-printed, and is shown in Figure \ref{fig:impressao-3d}. The concept behind using this sphere is that sensors placed inside it experience a wide range of accelerations and movements as it moves. By allowing free movement, the aim is to decouple the recorded signals from the typical motion biases of vehicles. The sphere has a diameter of \textbf{25 cm} and was carved from a parallelepiped measuring \textbf{25 cm x 18 cm x 10 cm}, with its center shifted by \textbf{5 cm} relative to the solid sphere, resulting in the model described here.

\begin{figure}[thb]
  \centering
  \includegraphics[width=1.0\linewidth]{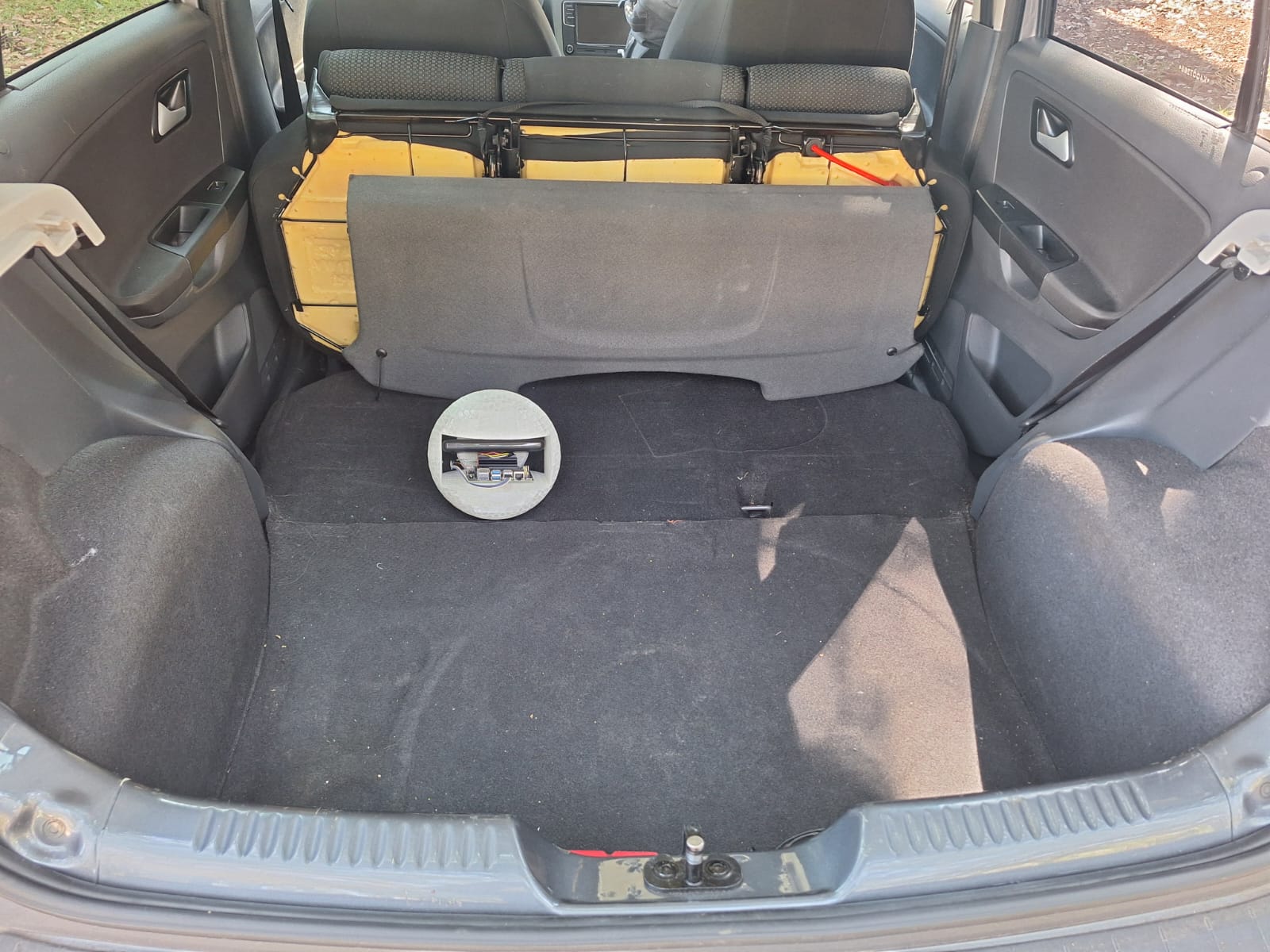}
  \caption{The sphere containing the entire collection set used for the dataset, within the vehicular platform. The structure moves freely during the journey.}
  \label{fig:carro}
\end{figure}

\begin{figure}[thb]
  \centering
  \includegraphics[width=1.0\linewidth]{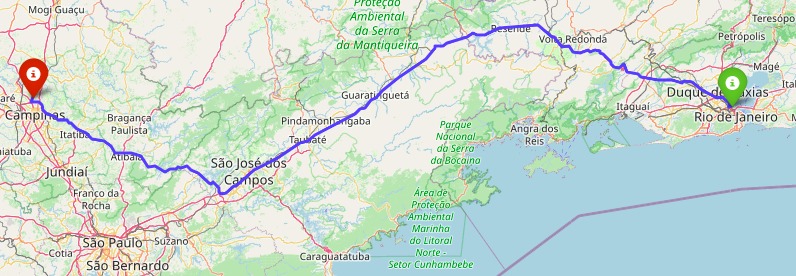}
  \caption{Trajectory followed for dataset collection. Source: OpenRouteService (HeiGIT) | Map data © OpenStreetMap contributors.}
  \label{fig:mapa}
\end{figure}

A moving platform is still necessary to set the sphere in motion and record different types of accelerations along its trajectory. For this purpose, a vehicle with an expanded trunk was used (Figure \ref{fig:carro}), travelling approximately 135 minutes (Figure \ref{fig:mapa}), between the Universidade Estadual de Campinas (UNICAMP, Zeferino Vaz campus) and the Federal University of Rio de Janeiro (UFRJ, Cidade Universitária campus). 
The final dataset is published as a CSV file with 10 columns: the first column corresponds to timestamps (in seconds), and the following nine columns represent the three axes of the accelerometer (in meters per second squared), gyroscope (in radians per second), and magnetometer (unit-normalized, pointing along the magnetic field direction). This dataset includes over 1.6 million consecutive samples (both raw and calibrated), a volume previously found only in datasets characterized by highly repetitive movement patterns (e.g., pedestrian locomotion) and/or low sampling rates.

\subsection{IMU Reference Frames and Calibration}
\label{sec:imu-calib}

The IMU provides measurements from an accelerometer, a gyroscope and a magnetometer, each expressed in the sensor's body frame. Interpreting these measurements in an inertial (world) frame requires knowledge of the sensor orientation and the removal of systematic sensor imperfections via calibration. The accelerometer measurement model is
\[
\mathbf{a}_m = \mathbf{S}_a (\mathbf{a}_{true} + \mathbf{g}) + \mathbf{b}_a + \mathbf{n}_a,
\]
where \(\mathbf{S}_a\) is the accelerometer scale matrix, \(\mathbf{b}_a\) the bias, \(\mathbf{n}_a\) measurement noise, and \(\mathbf{g}\) the gravity vector expressed in the sensor frame. The gyroscope is modeled as
\[
\boldsymbol{\omega}_m = \mathbf{S}_g \boldsymbol{\omega}_{true} + \mathbf{b}_g + \mathbf{n}_g,
\]
with \(\mathbf{S}_g\), \(\mathbf{b}_g\) and \(\mathbf{n}_g\) denoting scale, bias and noise respectively. The magnetometer, which is affected by hard-iron (additive) and soft-iron (multiplicative/shape) distortions, is represented by
\[
\mathbf{m}_m = \mathbf{S}_s (\mathbf{m}_{true} + \mathbf{b}_h) + \mathbf{n}_m,
\]
where \(\mathbf{b}_h\) denotes the hard-iron bias and \(\mathbf{S}_s\) encodes soft-iron effects.

To correct these systematic effects we applied standard procedures: the six-orientation method for the accelerometer, stationary bias estimation for the gyroscope, and ellipsoid fitting for the magnetometer. In the six-orientation procedure the IMU is placed in six known orthogonal poses to estimate \(\mathbf{b}_a\) and \(\mathbf{S}_a\) by enforcing
\[
\|\mathbf{a}_m - \mathbf{b}_a\| \approx g,
\]
with \(g\) the known gravitational acceleration. Gyroscope bias \(\mathbf{b}_g\) is estimated under static conditions (\(\boldsymbol{\omega}_{true}=\mathbf{0}\)) as the sample mean:
\[
\mathbf{b}_g = \frac{1}{N}\sum_{i=1}^{N} \boldsymbol{\omega}_m[i].
\]
For the magnetometer, we collected measurements across many orientations and performed ellipsoid fitting to estimate \(\mathbf{b}_h\) and \(\mathbf{S}_s\). The objective is to transform the measured ellipsoid into a unit sphere, i.e.
\[
(\mathbf{S}_s (\mathbf{m}_m - \mathbf{b}_h))^\top (\mathbf{S}_s (\mathbf{m}_m - \mathbf{b}_h)) = 1,
\]
thereby centering the cloud (hard-iron correction) and correcting its shape (soft-iron correction).

These calibrations remove offsets, scale factors and inter-axis misalignments, enabling consistent interpretation of sensor outputs in a common reference and preparing the data for sensor fusion and machine-learning pipelines. The full calibration sequence and scripts are provided in the dataset repository (section \ref{sec:data_available}).

\section{Results} 
\label{sec:results}

\begin{table}[thb]
\centering
\caption{Accelerometer calibration parameters}
\begin{tabular}{cccc}
\hline
Parameter & X-axis & Y-axis & Z-axis \\
\hline
Bias (m/s²) & 0.26784 & -0.21032 & 0.21283 \\
Scale & 1.00522 & 0.99968 & 1.01633 \\
Std Bias (m/s²) & 0.05613 & 0.05261 & 0.06113 \\
\hline
\end{tabular}
\label{tab:accel}
\end{table}

\begin{table}[thb]
\centering
\caption{Gyroscope calibration parameters}
\begin{tabular}{cccc}
\hline
Parameter & X-axis & Y-axis & Z-axis \\
\hline
Bias (rad/s) & 0.000978 & -0.004603 & 0.004887 \\
Std Bias (rad/s) & 0.001206 & 0.001148 & 0.001159 \\
\hline
\end{tabular}
\label{tab:gyro}
\end{table}

\begin{table}[thb]
\centering
\caption{Magnetometer calibration parameters (unit-normalized)}
\begin{tabular}{cccc}
\hline
Parameter & X-axis & Y-axis & Z-axis \\
\hline
Hard Iron Bias & -0.51625 & 0.32675 & 1.17227 \\
Scale & 1.00392 & 1.00564 & 1.00459 \\
\hline
\end{tabular}
\label{tab:magneto-bias}
\end{table}

\begin{table}[thb]
\centering
\caption{Magnetometer Soft Iron matrix}
\begin{tabular}{cccc}
\hline
Axis & X & Y & Z \\
\hline
X & 0.94163 & -0.07392 & 0.16855 \\
Y & -0.073923 & 0.88483 & 0.12147 \\
Z & 0.16855 & 0.12147 & 1.21098 \\
\hline
\end{tabular}
\label{tab:magneto-soft}
\end{table}

Table~\ref{tab:accel} shows the accelerometer calibration parameters. The bias values remain below $0.3\,m/s^2$, which is considered adequate for MEMS (\textit{Micro-Electro-Mechanical Systems}) sensors. The scale factors are very close to 1.0 (with variations below 2\%), confirming the good linearity of the sensor. The standard deviation of the bias (\textit{Std Bias}) ranges between $0.056\,m/s^2$ and $0.062\,m/s^2$ , within the expected range for this type of device.

In Table~\ref{tab:gyro}, corresponding to the gyroscope, the obtained biases are below $0.005\,rad/s$, which is approximately $0.25^{\circ}/s$, consistent with low-cost embedded inertial sensors. The standard deviations are also low (around $0.001\,rad/s$), indicating good stability.

Table~\ref{tab:magneto-bias} shows the magnetometer parameters. The hard iron bias values range from $-0.52$ to $+1.17$, which is acceptable and correctable during calibration, although the Z-axis is more affected. The scale factor is almost unity on all axes, consistent with the range of values provided by the hardware and predicted in its manual. The term \textit{hard iron} refers to magnetic distortions caused by permanent field sources near the sensor, such as magnets or DC currents, which shift the center of the measured ellipsoid.

\begin{figure}[bht]
  \centering
  \includegraphics[width=1.0\linewidth]{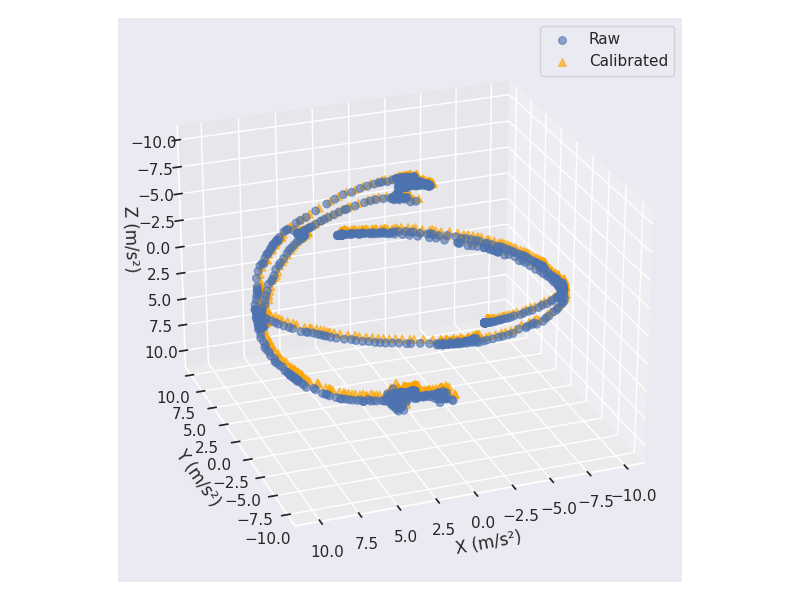}
  \caption{3D scatter plot of accelerometer measurements. Raw data (blue circles) and calibrated data (orange triangles) are shown.}
  \label{fig:accel_esfera}
\end{figure}

Table~\ref{tab:magneto-soft} presents the soft iron correction matrix. The diagonal elements differ from 1.0, ranging approximately from $0.88$ to $1.21$, indicating moderate distortions. Off-diagonal terms remain relatively small ($<0.2$), suggesting limited coupling between axes. The term \textit{soft iron} refers to distortions due to nearby ferromagnetic materials, which deform magnetic field lines, resulting in a tilted ellipsoid relative to the ideal axes.

\begin{figure}[bht]
  \centering
  \includegraphics[width=1.0\linewidth]{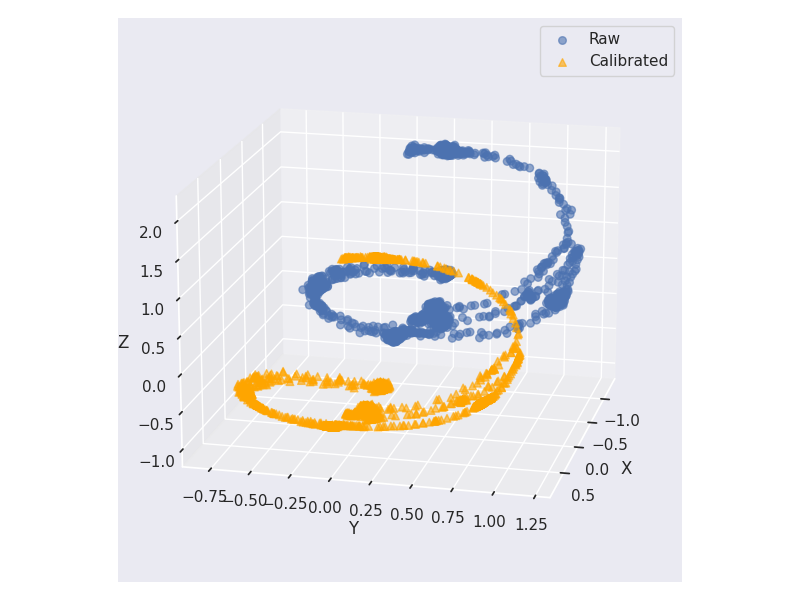}
  \caption{3D scatter plot of magnetometer measurements. Raw data (blue circles) and calibrated data (orange triangles) are shown.}
  \label{fig:mag_esfera}
\end{figure}

The estimated calibration parameters adjust the raw measurements, as can be visually verified. Scatter plots for sensors referenced to an inertial frame (accelerometer and magnetometer) were generated over a 2-minute segment of the dataset, comparing pre- and post-calibration measurements. The accelerometer primarily relies on the inertial frame defined by Earth's gravitational field, represented as a constant vector pointing toward the center of the planet. Therefore, after calibration, the plot in Figure \ref{fig:accel_esfera} shows trajectories centered around the origin with a radius of approximately $9.81\,m/s²$. The magnetometer relies on another inertial reference defined by Earth's magnetic field, pointing toward magnetic north. Prior to calibration, a significant bias is evident (as expected from Table \ref{tab:magneto-bias} and typical of uncalibrated magnetometers), but after calibration, the data dispersion (Figure \ref{fig:mag_esfera}) shows that measurements are consistently unit-length and centered around the origin. Corrections for scale and ellipsoid adjustment are minor, as expected from the calibration parameters, and visually imperceptible in the plots of Tables \ref{tab:accel} and \ref{tab:magneto-soft}.

Overall, the results demonstrate that all three sensors exhibit biases and non-linearities within typical limits for low-cost MEMS-based inertial units. These microfabricated sensors detect acceleration, angular velocity, and magnetic field intensity with small form factor, low power consumption, and affordability, albeit subject to noise and drift. The estimated corrections allow reliable use of the data in sensor fusion and machine learning applications, compatible with the type of signals encountered in such tasks. With the diversity of motions and high sampling rate enabled by the innovative dataset acquisition, the training of large-scale AI models in the inertial sensor domain is facilitated.

\section{Conclusion}
\label{sec:conclusion}

We presented a novel high-frequency (200Hz) inertial dataset obtained with a UM7 IMU placed inside a 3D-printed spherical structure, yielding long, diverse multichannel sequences with reduced motion bias. Calibration procedures for accelerometer, gyroscope, and magnetometer were applied and shown to place sensor errors within expected ranges for low-cost MEMS devices. Crucially, the use of a low-cost IMU and the natural noise introduced by recording inside a moving vehicle across many motion patterns is intended to enable models to learn and mitigate not only the underlying physical dynamics but also the characteristic measurement errors—analogous to how large language models learn phenomena and systematic artifacts from massive, unstandardized text corpora. By providing long-duration, high-resolution, and largely unlabeled inertial data, this resource facilitates self-supervised and Transformer-based approaches in the inertial domain. As immediate future work, we plan to develop and train a foundational model for inertial time series on the dataset released here, aiming to demonstrate improved generalization in tasks such as event detection, motion-mode recognition, and inertial navigation.

\section{Acknowledgements}
This work was partially funded by the Coordenação de Aperfei\c{c}oamento de Pessoal de Nível Superior – Brasil (CAPES) – Finance Code 001 and by the Ministry of Science, Technology, and Innovation of Brazil, with resources granted by the Federal Law 8.248 of October 23, 1991, under the PPI-Softex, DOU 01245.003479/2024-10.
The authors are also affiliated with the Artificial Intelligence Lab, Recod.ai.

\section{Compliance with Ethical Standards statement}
\label{sec:Compliance}

The data collection did not involve human subjects or require ethical approval.

\section{Data Availability}
\label{sec:data_available}

The IF-D dataset~\cite{ferreira_2025_17148626} is available in the link: \url{https://zenodo.org/records/17148626}.

\bibliographystyle{IEEEbib}
\bibliography{main}

\begin{thebibliography}{10}

\bibitem{ElSheimy2020}
Naser El-Sheimy and Ahmed Youssef,
\newblock ``{Inertial sensors technologies for navigation applications: state of the art and future trends},''
\newblock {\em Satellite Navigation}, vol. 1, no. 1, pp. 1--21, dec 2020.

\bibitem{Gao2022}
Yi~Gao, Fei Li, and Jia Chen,
\newblock ``{Random Weighting Adaptive Estimation of Model Errors on Attitude Measurement for Rotary Steerable System},''
\newblock {\em IEEE Access}, vol. 10, pp. 80794--80803, 2022.

\bibitem{Zhao2025TartanIMU}
Shibo Zhao, Sifan Zhou, Raphael Blanchard, Yuheng Qiu, Wenshan Wang, and Sebastian Scherer,
\newblock ``{Tartan IMU}: A light foundation model for inertial positioning in robotics,''
\newblock in {\em Proc. {IEEE/CVF} Conf. on Computer Vision and Pattern Recognition (CVPR)}, 2025.

\bibitem{Wieland2025}
Christoph Wieland and Victor Pankratius,
\newblock ``Inertial signal forecasting with foundation model techniques,''
\newblock {\em {IEEE} Sensors Journal}, 2025.

\bibitem{Zhou2021Informer}
Haoyi Zhou, Shanghang Zhang, Jieqi Peng, Shuai Zhang, Jianxin Li, Hui Xiong, and Wancai Zhang,
\newblock ``Informer: Beyond efficient transformer for long sequence time-series forecasting,''
\newblock May 2021, vol.~35, pp. 11106--11115.

\bibitem{Wu2021Autoformer}
Haixu Wu, Jiehui Xu, Jianmin Wang, and Mingsheng Long,
\newblock ``Autoformer: Decomposition transformers with auto-correlation for long-term series forecasting,''
\newblock in {\em Advances in Neural Information Processing Systems}, M.~Ranzato, A.~Beygelzimer, Y.~Dauphin, P.S. Liang, and J.~Wortman Vaughan, Eds. 2021, vol.~34, pp. 22419--22430, Curran Associates, Inc.

\bibitem{Zhou2022FEDformer}
Tian Zhou, Ziqing Ma, Qingsong Wen, Xue Wang, Liang Sun, and Rong Jin,
\newblock ``{FEDformer}: Frequency enhanced decomposed transformer for long-term series forecasting,''
\newblock in {\em International Conference on Machine Learning (ICML)}, 2022, pp. 27268--27286.

\bibitem{Nie2023PatchTST}
Yuqi Nie, Nam~H. Nguyen, Phanwadee Sinthong, and Jayant Kalagnanam,
\newblock ``A time series is worth 64 words: Long-term forecasting with transformers,''
\newblock in {\em Proc. Int. Conf. on Learning Representations (ICLR)}, 2023.

\bibitem{Lin2024SparseTSF}
Shengsheng Lin, Weiwei Lin, Wentai Wu, Haojun Chen, and Junjie Yang,
\newblock ``Sparsetsf: modeling long-term time series forecasting with 1k parameters,''
\newblock in {\em Proceedings of the 41st International Conference on Machine Learning}. 2024, ICML'24, JMLR.org.

\bibitem{Das2024TimesFM}
Abhimanyu Das, Weihao Kong, Rajat Sen, and Yichen Zhou,
\newblock ``A decoder-only foundation model for time-series forecasting,''
\newblock 2024.

\bibitem{Wu2023TimesNet}
Haixu Wu, Tengge Hu, Yong Liu, Hang Zhou, Jianmin Wang, and Mingsheng Long,
\newblock ``{TimesNet}: Temporal 2d-variation modeling for general time series analysis,''
\newblock in {\em Proc. Int. Conf. on Learning Representations (ICLR)}, 2023.

\bibitem{Anguita2013}
Jorge Reyes-Ortiz, Davide Anguita, Alessandro Ghio, Luca Oneto, and Xavier Parra,
\newblock ``Human activity recognition using smartphones,'' UCI Machine Learning Repository, 2013,
\newblock DOI: 10.24432/C54S4K.

\bibitem{Yan2020RoNIN}
Sachini Herath, Hang Yan, and Yasutaka Furukawa,
\newblock ``{RoNIN: Robust Neural Inertial Navigation in the Wild: Benchmark, Evaluations, New Methods},''
\newblock {\em Proceedings - IEEE International Conference on Robotics and Automation}, pp. 3146--3152, may 2020.

\bibitem{Burri2016EuRoC}
Michael Burri, Janosch Nikolic, Pascal Gohl, Thomas Schneider, Joern Rehder, Sammy Omari, Markus~W Achtelik, and Roland Siegwart,
\newblock ``The euroc micro aerial vehicle datasets,''
\newblock {\em Int. J. Rob. Res.}, vol. 35, no. 10, pp. 1157–1163, Sept. 2016.

\bibitem{Schubert2018TUMVI}
David Schubert, Thore Goll, Nikolaus Demmel, Vladyslav Usenko, Jörg Stückler, and Daniel Cremers,
\newblock ``The tum vi benchmark for evaluating visual-inertial odometry,''
\newblock in {\em 2018 IEEE/RSJ International Conference on Intelligent Robots and Systems (IROS)}, 2018, pp. 1680--1687.

\bibitem{Geiger2013KITTI}
A~Geiger, P~Lenz, C~Stiller, and R~Urtasun,
\newblock ``Vision meets robotics: The kitti dataset,''
\newblock {\em Int. J. Rob. Res.}, vol. 32, no. 11, pp. 1231–1237, Sept. 2013.

\bibitem{Shurin2022}
Artur Shurin, Alex Saraev, Mor Yona, Yevgeni Gutnik, Sharon Faber, Aviad Etzion, and Itzik Klein,
\newblock ``The autonomous platforms inertial dataset,''
\newblock {\em IEEE Access}, vol. 10, pp. 10191--10201, 2022.

\bibitem{ferreira_2025_17148626}
Patrick Ferreira and Paula Costa,
\newblock ``If-d: A high-frequency, general-purpose inertial foundation dataset for self-supervised learning,'' Sept. 2025.

\end{thebibliography}

\end{document}